# Simulation of Pedestrian Evacuation Using Conjugated Forces Model to Explore the Optimal Bypassing Strategy

Xiaolu Jia[1], Hao Yue[2], Dongliang He[3]

[1,2,3]MOE Key Laboratory for Urban Transportation Complex Systems Theory and Technology,
Beijing Jiaotong University, Beijing 100044, China
jiaxl@bjtu.edu.cn; yuehao@bjtu.edu.cn; 14120940@bjtu.edu.cn

***Abstract*-**The conjugated forces model (CFM) capable of reproducing bypassing behaviour is proposed and adopted to simulate pedestrian evacuation. Primarily, the concept of collision and evading and surpassing behaviour is particularly defined to describe the proposed CFM. A pedestrian whose behaviour can be described by the CFM will change his desired direction when confronted with a collision, and choose to actively bypass others. Secondly, pedestrian movement in a passageway is simulated to testify the capacity of the CFM to reproduce microscopic bypassing behaviour, and the bypassing parameter is then introduced to describe different movement strategies. Finally, Evacuation form a one-exit room is simulated to find the optimal bypassing strategy. It is concluded that the optimal bypassing strategy varies with different exit width and initial pedestrian number.

## 1. Introduction

Study on evacuation is of significance to the design of architecture layout, walking facilities, and evacuation schemes. However, empirical data is hard to achieve, and evacuation experiments are costly in experiment conduction and data analysis, especially when the layout of the walking facility changes. Models and simulation of evacuation is therefore widely accepted as a low-cost approach to study evacuation in real life. Pedestrian simulation models can be mainly classified as the macroscopic model [1] and microscopic model, and the latter one is more widely used in the state-of-the-art research.

Microscopic models mainly include the social force model (SFM) [2], centrifugal forces model [3], cellular automata (CA) model [4, 5] and combined models [6, 7]. The social force model proposed by Helbing et al. assumed that a pedestrian moved at the influence of driven force, repulsion force and friction, and successfully reproduced macroscopic behaviour like arching and "faster-is-slower effect" [2]. The centrifugal forces model by Yu et al. considered the effects of both the headway and the relative velocity among pedestrians, and reproduced the arching and the self-organization phenomenon [3]. Compared with the two continuous models introduced before, cellular automata model [4, 5] is superior for its high computation speed owing to its discreteness. Albeit the CA model can reproduce certain macroscopic phenomena, it is more artificial since it divides the space into discrete grids. To combine the advantages of continuous and discrete models, some models [6, 7] combine the SFM and floor field model, and successfully promote the computation speed and the authenticity of simulation results.

Based on the models introduced above, many researchers are concerned about the reproduction of microscopic behaviour, especially studies of bypassing behaviour, through improving the existing models. Seyfried proposed a one-dimensional social force model and set the velocity of the pedestrian who would bump into another at the next time step to be zero [8]. Parisi proposed that the desired velocity of a pedestrian will be set to zero if he detects obstacles in his respect area [9]. Rudloff et al. improved the repulsion force in the SFM by converting it to two orthogonal repulsion forces and calibrated relative parameters based on empirical data [10]. Frank reproduced the bypassing behaviour through changing the desired direction of a pedestrian when he was obstructed by an obstacle, and the improved desired direction pointed to the edge of the obstacle [11]. To test the capacity of the model to evade static obstacles in a relatively complex situation, many researchers took the classroom as the simulation scenario and simulated the evaluation process based on improved CA models [12, 13, 14]. Ming Tang constructed the bypassing behaviour model based on the velocity-time domain, and analysed the pedestrian arching at bottlenecks [15].

On the other hand, evacuation strategy is an important study subject in the implement of simulation models. Yue constructed a mixed exit selection strategy in pedestrian evacuation simulation with multi-exits, and found a critical density to distinguish the occurrence of pedestrian jam and whether an exit selection strategy is effective [16]. Yang and Dong modified the SFM to simulate guided crowd dynamics, and pedestrians in the guided crowd model adopted time-dependent desired velocities [17]. Hou and Liu assumed the desired velocity of a pedestrian pointed to the nearest guide, and then explored the optimal quantity and distribution of guides during evacuation through simulation [18].

In this work, it is assumed that pedestrians are homogeneous, and all the pedestrians share the same bypassing extent. The remainder of the paper is organized as follows: In section 2, the conjugated forces model is constructed based on the definition of collision and bypassing behaviour. In section 3, we validate the proposed CFM model through reproducing microscopic behaviour, introduce the bypassing parameter to describe different movement strategy, and eventually simulate the evacuation from a one-exit room to explore the optimal bypassing strategy.

## 2. Evacuation Model

### 2.1. Definition of collision

A pedestrian in a room is usually willing to move in his desired direction at a desired velocity, and this behaviour is defined as his desired movement status. However, the pedestrian cannot always fulfill his desired movement status since obstacles or other pedestrians may obstruct him from moving in the desired direction. Meanwhile, the pedestrian is physically unable to overlap with others, which means the pedestrian will collide with others when he is confronted with obstructions and simultaneously tries to keep his desired movement status. The concept of collision is explicitly defined for the calculation of conjugated forces.

It is assumed that $\boldsymbol{p_i}(t)$ is the position of pedestrian $i$ at time step $t$, and $\boldsymbol{v_i^0}(t)$ is the desired velocity of pedestrian $i$. The pedestrian $i$ at $\boldsymbol{p_i}(t)$ is assumed to be capable of predicting his future movement within a certain time period $t_i^d$, which can be defined as detection time period. If $i$ move at his desired velocity and will contact or overlap with pedestrian $j$ at time step $t^i$ while keeping clear of $j$ among time step $t$ and $t^i$, the time difference $t^i - t$ is defined as expected collision time $t_{ij}^c$. If pedestrian $i$ will never collide with $j$, $t_{ij}^c = \infty$. The distance between position $\boldsymbol{p_i}(t)$ and $\boldsymbol{p_i}(t')$ is defined as expected collision distance $d_{ij}^c$, and the relation between $t_{ij}^c$ and $d_{ij}^c$ is shown in Eqs. (1).

$$d_{ij}^c = \|\boldsymbol{p_i}(t) - \boldsymbol{p_i}(t')\| = \|\boldsymbol{v_i^0}(t)\| \cdot t_{ij}^c \tag{1}$$

The pedestrian $i$ will probably collide with multiple individuals, whereas he will change his movement status after colliding with other individuals for the first time. Therefore, the collision time $t_i^c$ and collision distance $d_i^c$ of pedestrian $i$ can be defined as follows:

$$t_i^c = \min_j t_{ij}^c, \ d_i^c = \min_j d_{ij}^c, j \in N, \ j \neq i, \tag{2}$$

Where $N$ is defined as the pedestrian set, and each pedestrian in the set has a collision time $t_i^c(t)$. Considering each pedestrian has a detection time period $t_i^d$, no collisions will be detected if $t_i^c(t) > t_i^d$. Therefore, if the collision time $t_i^c(t)$ of pedestrian $i$ is shorter that the detection time period $t_i^d$, pedestrian *i* is defined to have a collision with pedestrian *j* who will collide with *i* within the least expected collision time.

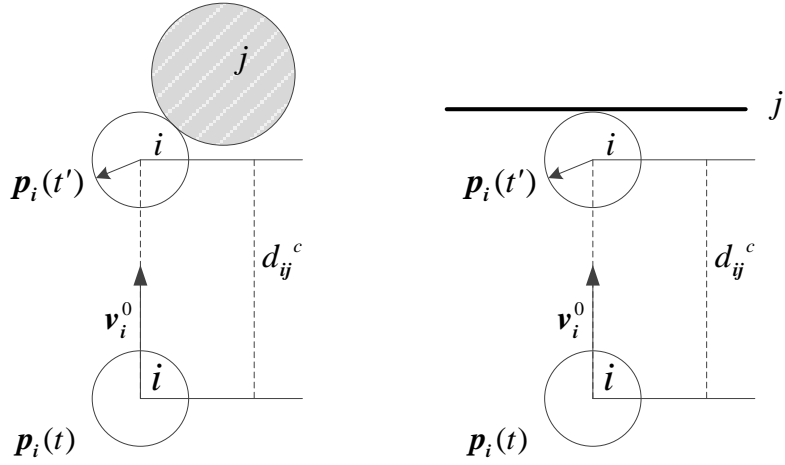

Fig. 1 Illustration of collision (a) with a circle-like obstacle (b) with a bar-like obstacle

## 2.2. Evading and surpassing behaviour

A pedestrian being about to collide with an obstacle will not remain the former desired movement status, i.e. desired velocity, but take actions to evade the obstacle. On the other hand, pedestrians will behave differently when having collisions with other pedestrians. Some people prefer to decelerate or stop to avoid colliding with others, while others choose to surpass for better movement status.

The pedestrian deciding to evade or surpass is assumed to keep the desired velocity magnitude constant and change the direction of the desired velocity to avoid collisions in the former desired direction. Conjugated Forces Model (CFM) is proposed to simulate the evading and surpassing behaviour. On the other hand, a pedestrian deciding not to surpass others will passively react to his local environments, which can be simulated by the Social Force Model (SFM).

## 2.3. Conjugated Forces Model

Conjugated Forces Model is constructed to change the desired directions or driven forces of pedestrians when they take evading or surpassing behaviour. The original definition of driven force $\boldsymbol{f}_i^0$ is as follows:

$$\boldsymbol{f}_i^0 = \boldsymbol{f}_i^+ - \boldsymbol{f}_i^-, \qquad (3)$$
$$\boldsymbol{f}_i^+ = m_i \boldsymbol{v}_i^0 / \tau, \qquad (4)$$
$$\boldsymbol{f}_i^- = m_i \boldsymbol{v}_i / \tau, \qquad (5)$$
$$\boldsymbol{v}_i^0 = \|\boldsymbol{v}_i^0\| \cdot \boldsymbol{e}_i, \qquad (6)$$

Where $m_i$ is the weight of pedestrian *i*, $\boldsymbol{e}_i$ is the desired direction, $\tau$ is the simulation time step, $\boldsymbol{f}_i^+$ is the desired movement status, $\boldsymbol{f}_i^-$ is the current movement status, and $\boldsymbol{f}_i^0$ is the motivation to convert from the current movement status to the desired status.

In the SFM, the desired movement status is always oriented to an exit, while in real life, some pedestrians will change the desired direction to find perceived better movement status when confronted with collisions. The CFM is therefore constructed to simulate the behaviour of those pedestrians. The total force of pedestrian *i* is as follows:

$$f_i = f_i^a + \sum_j f_{ij}^r, \tag{7}$$
$$f_i^a = f_i^s - f_i^-, \tag{8}$$

Where $f_i$ is the total force, $f_i^a$ is the driven force, $f_{ij}^r$ is the repulsion force, $f_i^s$ is the perceived desired movement status. It is noted that friction is not considered in this paper.

A pedestrian deciding to change the desired movement status will intend to decelerate in the former desired direction and accelerate in the tangential direction, thus changing the desired direction and driven force. Therefore, $f_i^c$ is defined as colliding force, which represents the driven force to decelerate in the former desired direction. $f_i^l$ is defined as changing force, which represents the driven force to accelerate in the tangential direction. The calculation of $f_i^s$, $f_i^c$ and $f_i^l$ is as follows:

$$f_i^s = f_i^+ + f_i^c + f_i^l, \tag{9}$$
$$f_i^c = A_i \exp(-t_i^c/B_i) \cdot (-e_i), \tag{10}$$
$$t_i^c = \max_{j \neq i} d_{ij}^c / (v_i^0 - v_j) \cdot e_i, \tag{11}$$
$$f_i^l = (2\|f_i^+\| \cdot \|f_i^c\| - \|f_i^c\|^2) \cdot d_i, \tag{12}$$

Where $A_i$ and $B_i$ are parameters representing the maximum magnitude of colliding force and reaction intensity respectively when pedestrian $i$ is confronted with a collision. It is defined that $A_i = \|f_i^a\|$, which insures the pedestrian would not retreat. Meanwhile, if $B_i \to 0^+$, pedestrian $i$ will not change his desired movement status; if $B_i \to \infty$, the desired direction will convert to its tangential direction. $d_i$ is a unit vector tangential to $e_i$. $f_i^l$ is calculated to keep the desired velocity constant, which can be illustrated in figure 2.

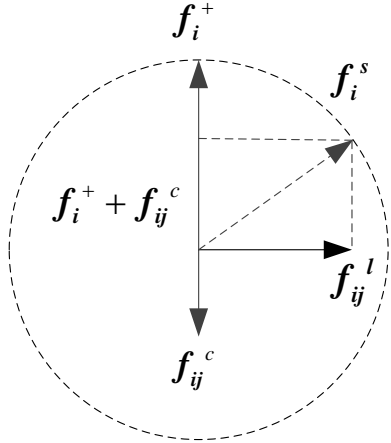 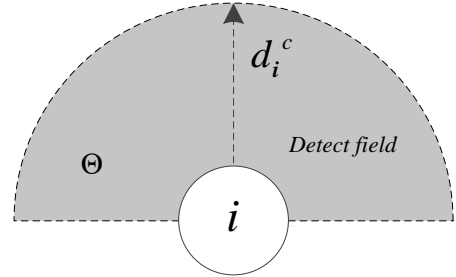

Fig. 2 Illustration of conjugated forces    Fig. 3 The detect field of pedestrian $i$

It can be seen that $f_i^+ + f_i^c$ and $f_i^l$ are a pair of conjugated forces, and jointly keep the desired velocity magnitude constant. The proposed model is therefore nominated as the Conjugated Force Model.

On the other hand, repulsion force $f_i^r$ represents the influence of pedestrian $j$ on $i$ when $j$ is in the detect field of $i$. The detect field is depicted in figure 3, which is set as a semicircle with the radius $d_i^c$.

$f_i^r$ points from $j$ to $i$ and reflects the intention of $j$ to keep away from $i$ when the two pedestrians do not contact, and represent the contact force when they contact with each other. The calculation of $f_i^r$ is as follows:

$$f_i^r = A_i \exp[-((d_{ij} - r_{ij}) + \|d_{ij} - r_{ij}\|)/(2 \cdot \|v_i^0\| \cdot B_i)] \cdot \mathbf{n}_{ij}, \tag{13}$$
$$d_{ij} = \|\mathbf{p}_i(t) - \mathbf{p}_j(t)\|, \tag{14}$$
$$r_{ij} = r_i + r_j, \tag{15}$$
$$\mathbf{n}_{ij} = (\mathbf{p}_i(t) - \mathbf{p}_j(t))/\|\mathbf{p}_i(t) - \mathbf{p}_j(t)\|, \tag{16}$$

Where $d_{ij}$ is the distance between pedestrian $i$ and $j$, $r_i$ is the radius of pedestrian $i$, and $\mathbf{n}_{ij}$ is a unit vector pointing from pedestrian $j$ to $i$.

## 3. Simulation and results

Pedestrian microscopicbehaviour in a passageway and evacuation in a room with one exit is simulated based on the proposed CFM. The values of relative parameters, which are not relevant to the simulation scenario,are shown in table1.

Table1. Values of simulation parameters

| parameter | definition | value | unit |
|---|---|---|---|
| $\tau$ | Simulation time step | 0.5 | s |
| $m_i$ | Weight of pedestrian $i$ | 60 | kg |
| $r_i$ | radius of pedestrian $i$ | 0.25 | m |
| $v_i^0$ | Desired velocity of pedestrian $i$ | 1.0 | m/s |
| $B_i$ | Collision parameter | 0.8 | m |
| $t_i^d$ | detection time period | 3 | s |

Based on the CFM model, pedestrian movement in a passageway is simulated to test the capacity of the proposed model to reproduce certain microscopic behaviour. Then, the bypassing parameter is introduced to represent different movement strategy. Finally, evacuation from a one-exit room is simulated to explore the optimal pedestrian movement strategy during evacuation.

### 3.1. Reproduction of microscopic behaviour

Microscopic pedestrian movement in a passageway (10m×5m) is simulated, and the movement trials of the individuals are shown in Fig 4 and Fig 5.Taking the pedestrian initially on the left side of the passageway as a study object, his movement trial of evading a circle-like obstacle and a bar-like obstacle is illustrated in Fig 4(a) and Fig 4(b), which proves the capability of the CFM to evade spatial obstacles. From Fig 4(c) and Fig 4(d), the pedestrian can also be seen that he is also to be capable of evading a pedestrian moving in the opposite direction, and surpassing a pedestrian in the same movement direction but at a lower velocity.Accordingly, it is illustrated that the pedestrian will positively bypass the pedestrian being in a collision with him.

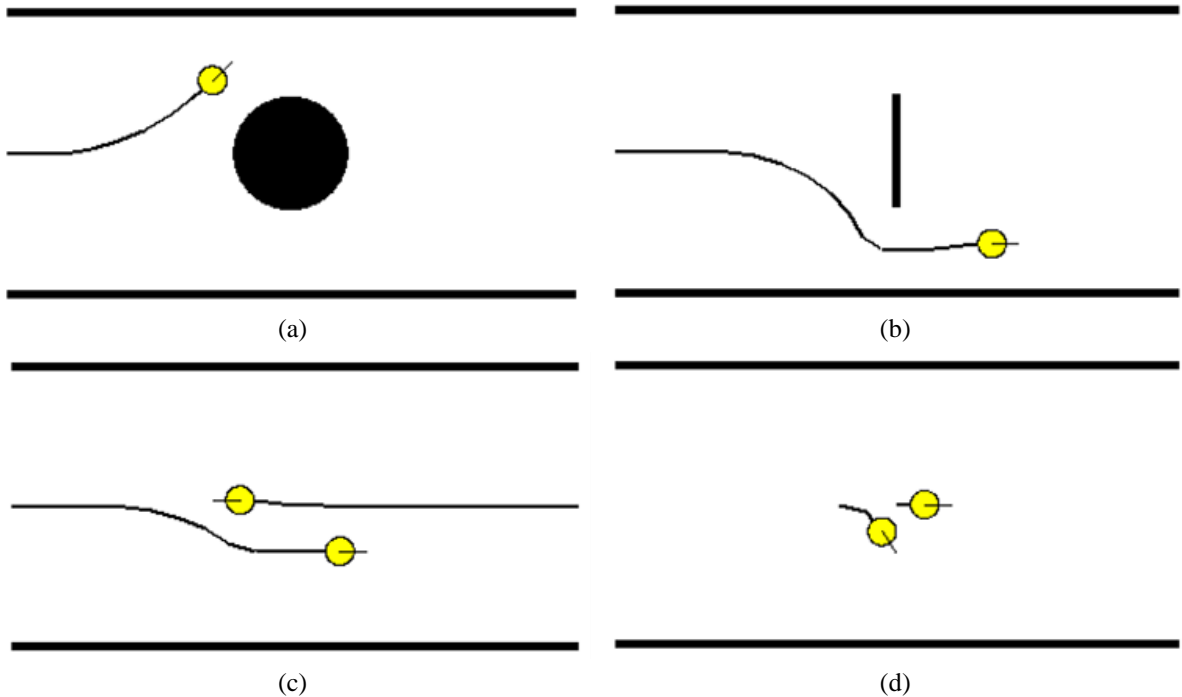

Fig. 4 Bypassing behaviour when facing with different obstacles

On the other hand, the CFM can reproduce different bypassing behaviour of a pedestrian under the same scenario owing to stochastic serial update mechanism. Taking the pedestrian initially on the right left side of the passageway in Fig. 5 as the study object, it can be seen that he keep his movement status in (a), evade the three opposite pedestrians as a whole in (b), and passes across those opposite pedestrians one-by-one in (c) and (d).

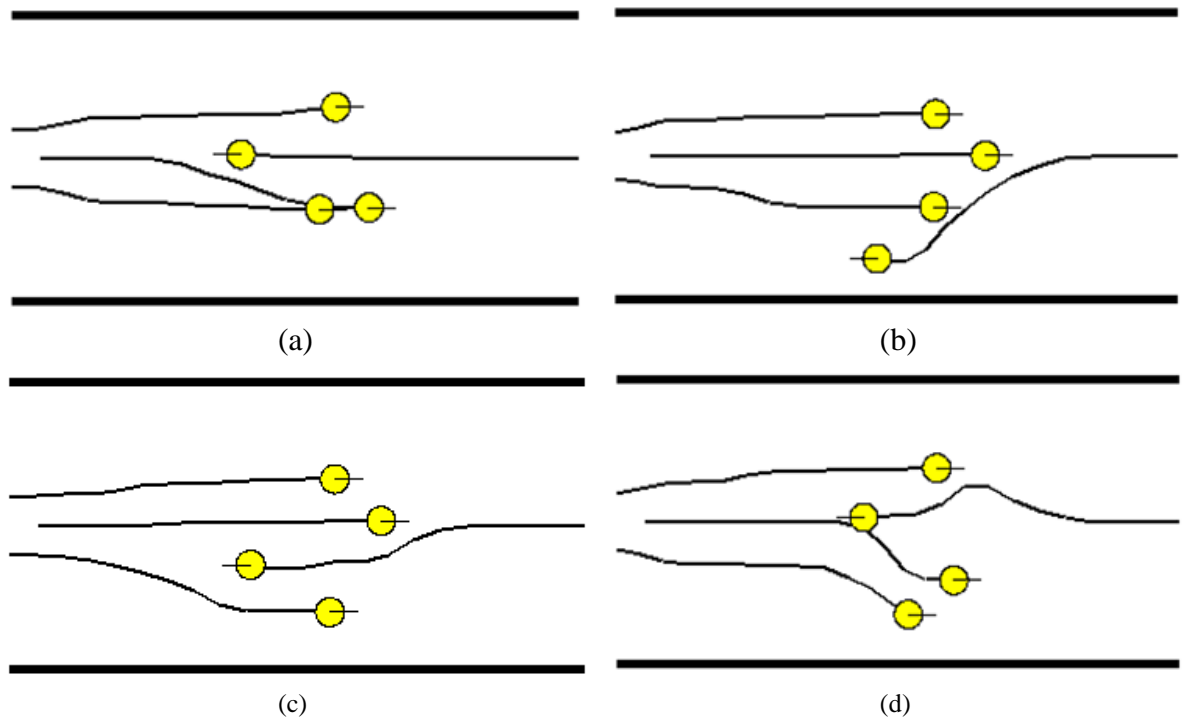

Fig. 5 Bypassing behaviour when facing multiple pedestrians

### 3.2. Bypassing parameter

Pedestrians adopting the behaviour in the SFM could be considered as passive ones, while those adopting the behaviour in the CFM could be considered as aggressive ones. Nevertheless, a pedestrian in the CFM will change his desired direction as soon as he is encountered with a collision, which means the pedestrian is oversensitive. While a pedestrian in the SFM will never bypass the pedestrian directly in front of him, which means the pedestrian is over-conservative.

To simulate the behaviour of normal pedestrians, the bypassing parameter $\alpha$ is introduced to redefine the collision time, which means the pedestrian will not change his desired direction until $t_i^c \leq \alpha \cdot t_i^d$. Different values of $\alpha$ can represent different pedestrian movement strategies. The pedestrian follows the SFM when $\alpha = 0$, while follows the CFM when $\alpha = 1$. When $0 < \alpha < 1$, the pedestrian will not immediately change his desired direction when he detects a collision, but wait until $t_i^c$ is not larger than $\alpha \cdot t_i^d$.

### 3.3. Optimal movement strategy of evacuation

Pedestrian evacuation in a room (15m×15m) without internal obstacles is simulated based on the proposed model. The initial pedestrian number, exit width, bypassing parameter and evacuation time are represented by $N$, $r$, $\alpha$ and $T$ respectively. The simulation scenario when $N = 60$ and $r = 1$ can be indicated in Fig 6, and the initial desired direction is directed to the centre of the exit.

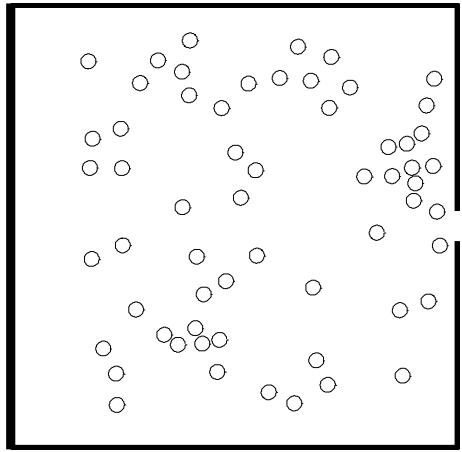

Fig. 6 Simulation scenario with $N = 60$ and $r = 1$.

Evacuation time with different exit width (r= 1, 3, 5) can be achieved through averaging the simulation results of 20 runs, which is illustrated in Fig 7. It can be indicated that the evacuation time is relevant to the exit width, initial pedestrian number and bypassing parameter.

Through contrasting the three subfigures in Fig 7, it is clear that $T$ decreases at the rise of $r$, and the difference value between different $T$ with different $r$ increases with the increase of $N$. This result is reasonable. When $r$ is small, the exit passing capacity is relatively lower because of the smaller space for pedestrians to evacuate, and the evacuation time is therefore larger. The evacuation space for pedestrians will further diminish with the increase of initial pedestrian number, and therefore enhances the evacuation time. However, the influence of bypassing parameter is not consistent under different situations.

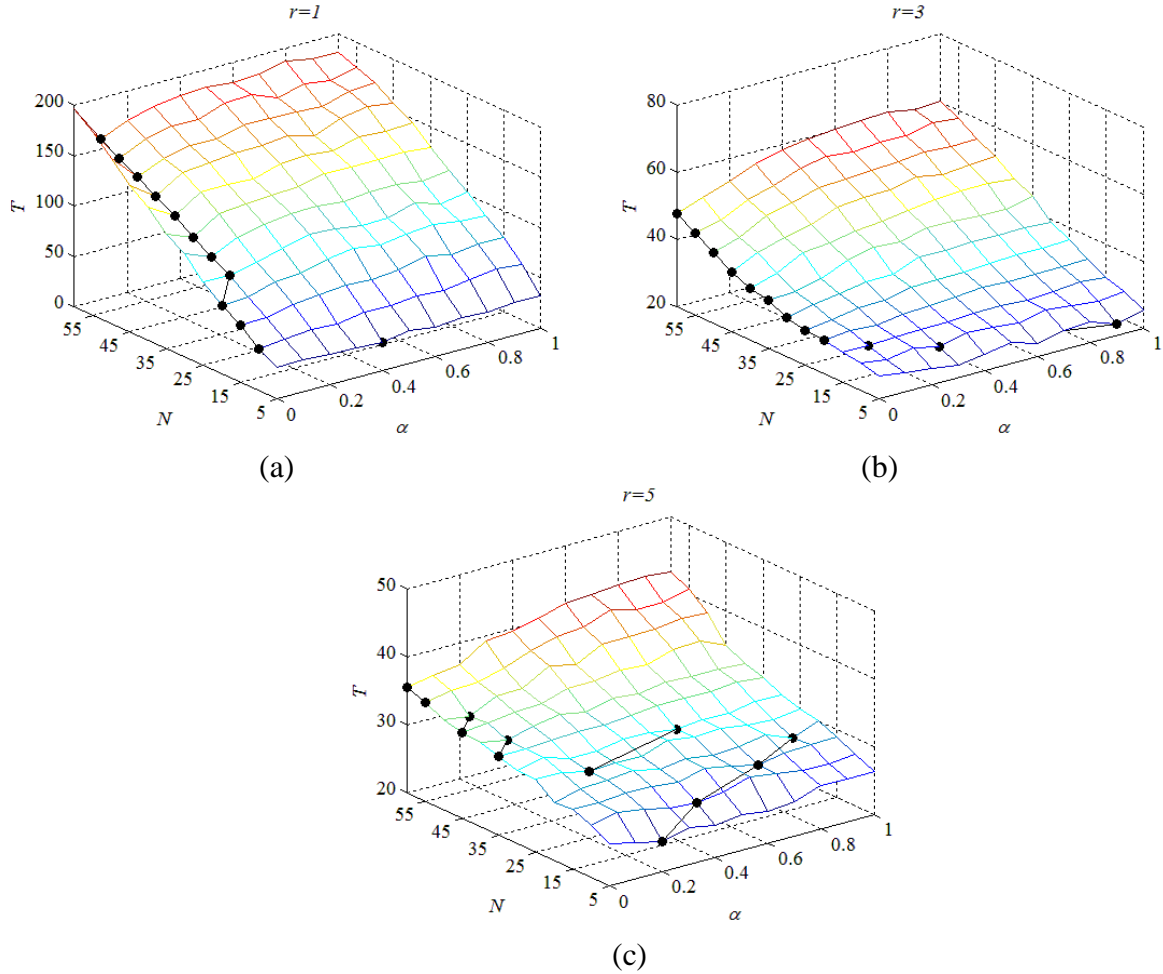

Fig. 7 The influence of $N$ and $\alpha$ on $T$ with different $r$.

    The optimal $\alpha$ under different $N$ is the one corresponding to the smallest $T$, and is represented by black dots in Fig 7. When $r=1$, the exit passing capacity is small, and the best movement strategy is not or slightly evade other pedestrians. That is to say, when the exit width is small, evacuating orderly will be the best choice. When $r=3$, the exit passing capacity is relatively larger than before, and the best strategy correlates to the value of $N$. When $N$ is large($N \geq 25$), it is wise for pedestrians to be passive, while when $N$ is small($N \leq 20$), adopting aggressive movement will bring about shorter evacuation time. When $r=5$, the exit passing capacity is relatively sufficient, and adopting a $\alpha$ is the best strategy in spite of when $N$ is too large($N \geq 45$).

    It is noted that the optimal $\alpha$ seems to be irregular when $r=5$, which is supposed to be the influence of initialization. The exit passing capacity under that situation is relatively sufficient, and pedestrians will unlikely to be congested before the exit, which means the initial position of pedestrians has a big influence on the evacuation time. Nevertheless, it can be concluded that pedestrians will attain shorter evacuation time through bypassing others when the exit width is larger and the initial pedestrian number is relatively smaller.

## 4. Conclusion

    The CFM is proposed and adopted to simulate pedestrian evacuation from a one-exit room, which contributes to the exploration of the optimal movement strategy during evacuation. The main propositions of this paper are as follows.

Primarily, the conjugated forces model (CFM) is constructed to reproduce the bypassing behaviour. A pedestrian whose behaviour can be described by the CFM will change his desired direction when confronted with collision, and choose to actively evade static obstacles and surpass mobile obstacles.

Secondly, pedestrian movement in a passageway is simulated to testify the capacity of the CFM to reproduce the bypassing behaviour. It is noted that pedestrians in the CFM can take different behaviour under the same simulation scenario, which is owing to the stochastic serial update mechanism.

Finally, to make the CFM model closer to reality, the bypassing parameter is introduced to describe the behaviour of strategic pedestrians. The value of bypassing parameter will rise with the increase of bypassing extent. Evacuation form a one-exit room is simulated to find the optimal bypassing parameter under each simulation scenario with certain exit width and initial pedestrian number.

It can be concluded that the optimal bypassing parameter, i.e. the optimal movement strategy, varies under different simulation scenarios. When the exit width is small, it is optimal for pedestrians to move orderly or bypass slightly. When the exit width is big enough, it is optimal to bypass others, and the optimal bypassing parameter seems irregular owing to the influence of initialization. When the exit width is moderate, it is optimal for pedestrians to move in order when the initial pedestrian number is above than the critical pedestrian number. Whereas when the initial pedestrian number is under the critical one, it is better to adopt bypassing behaviour, and the optimal bypassing parameter increases with the decrease of initial pedestrian number.

The CFM proposed in this paper can be implemented into the simulation of pedestrian evacuation, and the results can offer inspiration for the management of pedestrians during evacuation.

## Acknowledgements

Project supported by the Fundamental Research Funds for the Central Universities of China (Grant No. 2015JBM045), the National Science Foundation of China (Grant No. 51338008, 11172035,51308038), the National Basic Research Program of China (Grant No. 2012CB725403) and the Center of Cooperative Innovation for Beijing Metropolitan Transportation.

## References

[1] R. L. Hughes. "The flow of large crowds of pedestrians," Mathematics & Computers in Simulation, vol. 53, no. 4-6, pp. 367-370, 2000.
[2] D. Helbing, I. Farkas and T. Vicsek, "Simulating dynamical features of escape panic," *Nature*, vol.407, no. 6803, pp. 487-490, 2000.
[3] W. J. Yu, R. Chen, L. Y. Dong, and S. Q. Dai, "Centrifugal force model for pedestrian dynamics," *Phys.Rev.E*, no.72, no. 026112, 2005.
[4] V. Blue and J. Adler, "Modelling four-directional pedestrian flows," *Transp. Res. Rec*, 2000, vol. 1710, no. 1, pp. 20-27.
[5] C. Burstedde, K. Klauck, A. Schadschneider and J. Zittartz, "Simulation of pedestrian dynamics using a two-dimensional cellular automaton," *Physica A*, vol. 295, no. 3, pp. 507-525, 2001.
[6] W. G. Song, Y. F. Yu, B. H. Wang and W. C. Fan, "Evacuation behaviours at exit in CA model with force essentials: A comparison with social force model," *Physica A*, vol. 371, no. 2, pp. 658-666, 2006.
[7] R. Chen, X. Li and L. Y. Dong, "Modeling and simulation of weaving pedestrian flow in subway stations", *Acta Phys. Sin.*, vol. 61, no. 14, pp. 144502, 2012.
[8] A. Seyfried, B. Steffen and T. Lippert, "Basics of modelling the pedestrian flow," *Physica A*, vol. 368, no. 1, pp. 232-238, 2005.
[9] D. R. Parisi, M. Gilman and H Moldovan, "A modification of the Social Force Model can reproduce experimental data of pedestrian flows in normal conditions," *Physica A*, vol. 388, no. 17, pp. 3600-3608, 2009.
[10] C. Rudloff, T. Matyus, S. Seer and D. Bauer, "Can Walking Behaviour Be Predicted? An analysis of the calibration and fit of pedestrian models", *Transp. Res. Rec.*, vol. 2264, pp. 101-109, 2011.

[11] G. A. Frank and C. O. Dorso, "Room evacuation in the presence of an obstacle," *Physica A*, vol. 390, pp. 2135-2145, 2011.

[12] L. Chen, R. Y. Guo and N. Ta, "Simulation and experimental results of evacuation of pedestrian flow in a classroom with two exits," *Acta Phys. Sin.*, vol. 62, no. 5, pp. 050506, 2013.

[13] K. J. Zhu and L. Z. Yang, "The effects of exit position and internal layout of classroom on evacuation efficiency," *Acta Phys. Sin.*, vol. 59, no. 11, pp. 7701-07.

[14] L. Y. Dong, L. Chen and X. Y. Duan, "Modeling and simulation of pedestrian evacuation from a single-exit classroom based on experimental features," *Acta Phys. Sin.*, vol. 64, no. 22, pp. 220505, 2015.

[15] M. Tang, H. Jia, B. Ran and J. Li, "Analysis of the pedestrian arching at bottleneck based on a bypassing behaviour model," *Physica A*, vol. 453, pp. 242-258.

[16] H. Yue, B. Y. Zhang, C. F. Shao and Y. Xing, "Exit selection strategy in pedestrian evacuation simulation with multi-exits," *Chin. Phys. B*, vol. 23, no. 5, pp. 050512, 2014.

[17] X. X. Yang, H. R. Dong, Q. L. Wang, Y. Chen and X. M. Hu, "Guided crowd dynamics via modified social force model," *Physica A*, vol. 411, no 63, pp. 63-73, 2014.

[18] L. Hou, J. G. Liu, X. Pan and B. H. Wang, "A social force evacuation model with the leadership effect," *Physica A*, vol. 400, no. 2, pp. 93-99, 2014.